\documentclass{article}
\usepackage{spconf,amsmath,graphicx}
\usepackage{bbding} 
\usepackage{caption}
\usepackage{subcaption}
\usepackage{multirow}
\usepackage{url}
\usepackage{}
\usepackage{xcolor}
\usepackage{booktabs}
\usepackage{hyperref}
\hypersetup{hidelinks}
\newcommand*{\email}[1]{%
    \normalsize\href{mailto:#1}{#1}\par
    }

\title{MULTI-SPEAKER MULTI-STYLE TEXT-TO-SPEECH SYNTHESIS WITH SINGLE-SPEAKER SINGLE-STYLE TRAINING DATA SCENARIOS}
\name{Qicong Xie$^{1}$, Tao Li$^{1}$, Xinsheng Wang$^{1,2}$, Zhichao Wang$^{1}$, Lei Xie$^{1*}$ \thanks{$^{*}$Corresponding author.},  Guoqiao Yu$^{3}$, Guanglu Wan$^{3}$ }

\address{$^{1}$Audio, Speech and Language Processing Group (ASLP@NPU), School of Computer Science,\\ Northwestern Polytechnical University, Xi’an, China \\
$^{2}$School of Software Engineering, Xi’an Jiaotong University, Xi'an, China \\
$^{3}$Meituan-Dianping Group, Beijing, China \\
\email{xieqicong@mail.nwpu.edu.cn},~\email{taoli@npu-aslp.org},~\email{wangxinsheng@stu.xjtu.edu.cn},~\email{zcwang\_aslp@mail.nwpu.edu.cn},~\email{lxie@nwpu.edu.cn} }
\begin{document}
\ninept
\maketitle
\begin{abstract}
In the existing cross-speaker style transfer task, a source speaker with multi-style recordings is necessary to provide the style for a target speaker. However, it is hard for one speaker to express all expected styles. In this paper, a more general task, which is to produce expressive speech by combining any styles and timbres from a multi-speaker corpus in which each speaker has a unique style, is proposed. To realize this task, a novel method is proposed. This method is a Tacotron2-based framework but with a fine-grained text-based prosody predicting module and a speaker identity controller. Experiments demonstrate that the proposed method can successfully express a style of one speaker with the timber of another speaker bypassing the dependency on a single speaker's multi-style corpus. Moreover, the explicit prosody features used in the prosody predicting module can increase the diversity of synthetic speech by adjusting the value of prosody features.
\end{abstract}
\vspace{-4pt}
\section{Introduction}
\label{sec:intro}
\vspace{-6pt}

In recent years, enormous progress has been made in the neural text-to-speech (TTS), which benefits from the development of sequence-to-sequence (seq2seq) neural models~\cite{bahdanau2014neural,sutskever2014sequence}, making it possible to synthesize highly intelligible and natural speech~\cite{wang2017tacotron,shen2018natural,ren2019fastspeech,yu2020durian}. Despite the successful application of TTS in many scenarios, how to create expressive synthetic speech that can be flexibly controlled in terms of various speaking styles and speaker timbres is desirable for better user experience. This paper proposes a new expressive speech synthesis task that creates diversity synthetic speech by combining the timbre and speaking style from different speakers.

To create a TTS system with the ability to synthesize various expressive speech, a straightforward method is to train a TTS model with a database with manual labels~\cite{lee2017emotional,choi2019multi,litao2021controllable,li2018emphatic,liu2021expressive}, for instance, a database with manually labeled emotion categories~\cite{lee2017emotional,litao2021controllable} or speaking styles~\cite{liu2021expressive}. However, the limitation of these methods is obvious, i.e., it heavily depends on the training data and can not create new voice by combining different speaker timbres and speaking styles. To transplant a style to a target speaker for whom no labeled expressive recording exists, the cross-speaker style transfer task has attracted much attention~\cite{bian2019multi,whitehill2019multi,karlapati2020copycat,litao2021controllable,pan2021cross,shang2021incorporatingM3}. Reference embedding-based cross-speaker style transfer models~\cite{bian2019multi,whitehill2019multi,karlapati2020copycat,shang2021incorporatingM3,Li2021ControllableCE}, typically based on several general reference embedding methods \cite{skerry2018towards,wang2018style,zhang2019learning}, have shown promising performance on the style transfer task.

While those cross-speaker transfer methods can successfully produce expressive speech with a specific speaking style and a timbre from a speaker who has no such a speaking style in the corpus, they typically depend on a source speaker who has enough manually labeled expressive sources. It requires a source speaker to be an expert in expressing all expected styles with the aim to produce synthetic speech with various styles. Anyway, it is impossible for one source speaker to imitate all possible speaking styles and record enough recordings. In contrast, it is much easier to obtain an expressive corpus in which each speaker only speak one specific speaking style that he or
she is good at. With such a corpus, a practical task is to build a TTS system that has the ability to produce synthetic speech by combining different timbres and styles from different speakers, which is referred to as \emph{speaker-related multi-style and multi-speaker TTS (SRM2TTS)}. 

However, it is non-trivial to achieve such a SRM2TTS task. Compared with the traditional cross-speaker style transfer task, in the SRM2TTS task, the timbre and  style are closely entangled, making it difficult to transfer styles across speakers with reference-based methods. Taking inspiration from the success of the label-assisted content-aware prosody prediction model on the style transfer task~\cite{pan2021cross}, a novel method for the SRM2TTS task is proposed in this work. Specifically, based on a typical neural seq2seq framework, a content-aware multi-scale prosody modeling module is proposed, which can provide the style information to the TTS system based on the style label and input text. With an extra speaker identity controller, the proposed method can distinguish different styles and timbres, and thus can perform any combination of speakers and styles for SRM2TTS. Experiments have shown that the proposed method achieves good performance on synthesizing expressive speech by combining any speaker timbre and speaking style. Besides, benefit from the explicit modeling of prosody features, the proposed method can flexibly control each prosodic component, e.g., pitch and energy, which can increase the diversity of synthesized speech. 

\begin{figure*}[t]
    \centering
    \includegraphics[scale=0.5]{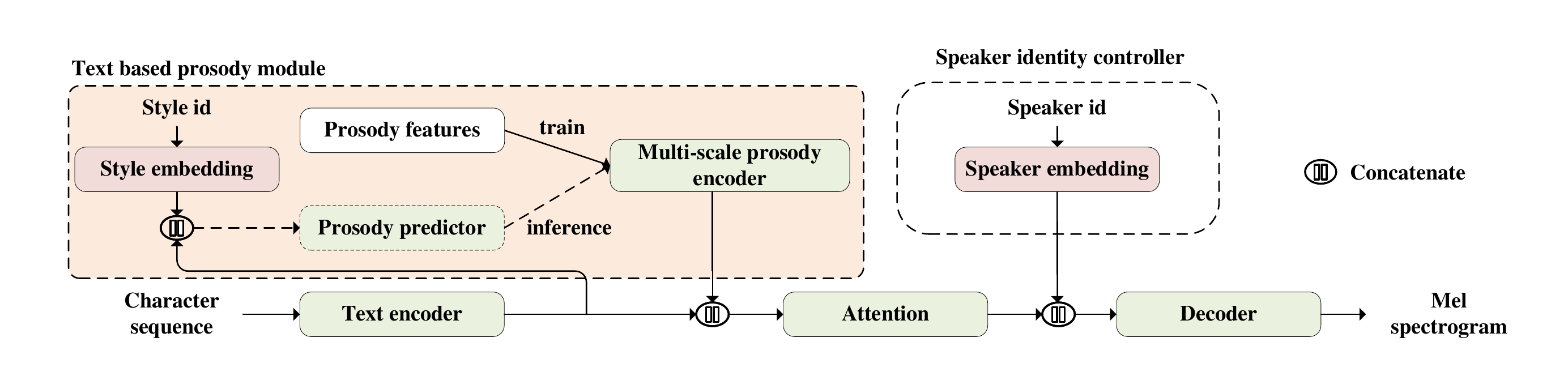}
    \caption{The architecture of the proposed model}
    \label{fig:model} \vspace{-0.4cm}
\end{figure*} 

\begin{figure}[t]
    \centering
    \includegraphics[scale=0.4]{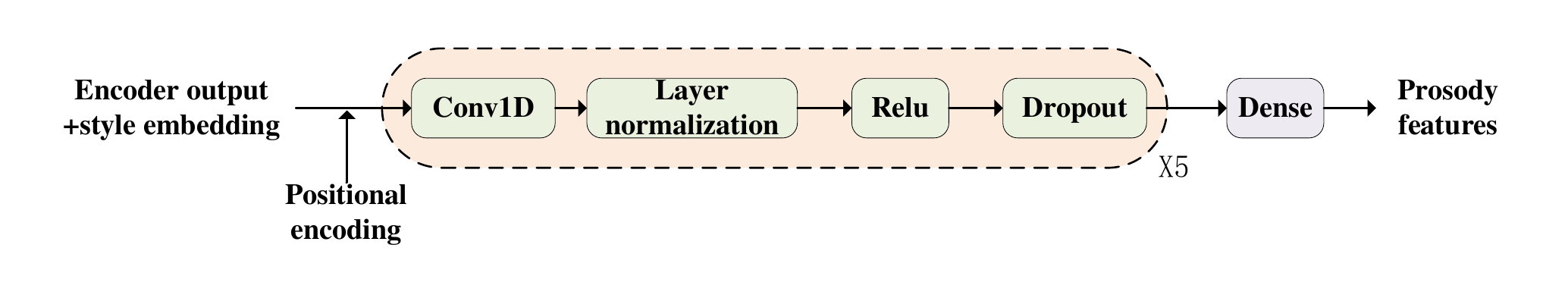}
    \caption{The architecture of the prosody predictor}
    \label{fig:pred}  
\end{figure} 

\begin{figure}[t]
    \centering
    \includegraphics[scale=0.45]{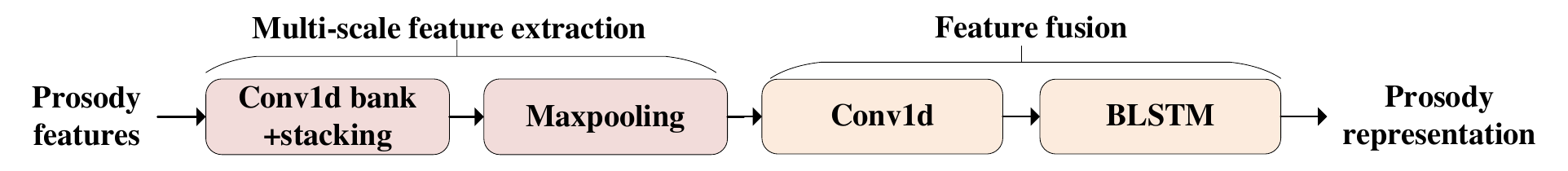}
    \caption{The architecture of the prosody encoder}
    \label{fig:enc} 
\end{figure} 

Our contribution can be summarized as follows: 
(1) Synthesizing expressive speech by combining any style and timbre based on a multi-speaker database in which each speaker has a unique speaking style is first proposed in this work. The realization of this task has profound implications from a perspective of usability.
(2) A novel method, which can realize the combination and control of any style and timbre on expressive speech synthesis, is proposed. 
(3) Extensive experiments have shown that with a novel fine-grained text-based prosody modeling module the proposed method can explicitly model and flexibly control the prosodic components.

\section{Proposed Model}

The proposed framework is illustrated in Fig.~\ref{fig:model}. As shown, the proposed model is a typical attention-based seq2seq framework, in which the backbone of the encoder-decoder structure is based on a modified Tacotron2~\cite{shen2018natural}. Besides, a novel text-based fine-grained prosody module is included to predict the prosody, and the speaker identity controller is to control the timbre of synthetic speech.

\subsection{The backbone of the encoder-decoder framework}

Following~\cite{litao2021controllable}, a slightly modified version of Tacotron2~\cite{shen2018natural} is used as the encoder-decoder backbone. The encoder consists of a pre-net, which is composed of two fully connected layers, and a CBHG module~\cite{lee2017fully}.
The decoder is composed of an autoregressive recurrent neural network (RNN) and generates attention queries at each decoder time step. Here, the GMM attention mechanism is used, which has shown a good performance on modeling long sequence speech~\cite{graves2013generating,battenberg2020location}. To control the timbre of synthesized speech, an additional speaker embedding with dimension of 256 is concatenated with the RNN input in the decoder. Same to the original Tacotron2~\cite{shen2018natural}, the post-net which is a five-layer convolution network is also adopted. Speech in this framework is represented by mel-spectrograms, and multi-band WaveRNN~\cite{yu2020durian} is adopted to reconstruct waveforms from predicted spectrograms.

\begin{table*}[!htbp]
 \caption{Comparison of our proposed method with Multi-R and PB in terms of style and speaker similarity MOS with confidence intervals of 95$\%$. The higher value means better performance, and the bold indicates the best performance out of three models in terms of each style.}
 \label{tab:transfer}
\setlength{\tabcolsep}{3mm}
 \centering
\begin{tabular}{l|lll|lll}
\toprule
\multicolumn{1}{c|}{\multirow{2}{*}{Style}} & \multicolumn{3}{c|}{style similarity MOS}                                                  & \multicolumn{3}{c}{speaker similarity MOS}                                                \\ \cmidrule{2-7} 
\multicolumn{1}{c|}{}                         & \multicolumn{1}{c}{Multi-R~\cite{bian2019multi}} & \multicolumn{1}{c}{PB~\cite{pan2021cross}} & \multicolumn{1}{c|}{Proposed} & \multicolumn{1}{c}{Multi-R~\cite{bian2019multi}} & \multicolumn{1}{c}{PB~\cite{pan2021cross}} & \multicolumn{1}{c}{Proposed} \\ \midrule
Story             & 3.59$\pm$0.058  & 3.65$\pm$0.056 & \bf{3.77}$\pm$\bf{0.058} & 3.83$\pm$0.049  & 3.91$\pm$0.049 & \bf{4.01}$\pm$\bf{0.047} \\
Anchor            & 3.48$\pm$0.061   & 3.61$\pm$0.057 & \bf{3.79}$\pm$\bf{0.058} & 3.74$\pm$0.050    & 4.01$\pm$0.047 & \bf{4.04}$\pm$\bf{0.048} \\
CS               & 3.22$\pm$0.060  & 3.78$\pm$0.062 & \bf{3.84}$\pm$\bf{0.059} & 3.81$\pm$0.046  & \bf{3.84}$\pm$\bf{0.048} & \bf{3.84}$\pm$\bf{0.046} \\
Poetry            & 2.84$\pm$0.057 & 3.88$\pm$0.060 & \bf{4.14}$\pm$\bf{0.054}  & 3.82$\pm$0.049 & \bf{3.88}$\pm$\bf{0.050} & 3.86$\pm$0.047 \\
Game              & 2.78$\pm$0.054 & 3.81$\pm$0.060 & \bf{4.03}$\pm$\bf{0.059} & 3.90$\pm$0.045 & 3.92$\pm$0.049 & \bf{4.04}$\pm$\bf{0.048}  \\ \midrule
Overall           & 3.18$\pm$0.023 & 3.74$\pm$0.027 & \bf{3.91}$\pm$\bf{0.026} & 3.82$\pm$0.021 & 3.91$\pm$0.022 & \bf{3.96}$\pm$\bf{0.021} \\ \bottomrule
\end{tabular}
\end{table*}

\subsection{Text-based fine-grained prosody module}

As mentioned in the introduction, when there is an exact correspondence between the speaking style and speaker identity, the speaker information and style information would be deeply entangled from a global perspective. Therefore, it is crucial to find the essential difference between the speaker information and style information. Actually, the speaker information, typically the timbre, is global information, which means the timbre related to speaker identity basically will not change along with the speaking style. In contrast, the speaking style, which is generally presented by  fine-grained prosody, is mainly local information and will varies with different speech units. Direct presenting the prosody as a global embedding is hard to distinguish from the speaker embedding in our case. Instead, a fine-grained prosody encoder, as shown in Fig.~\ref{fig:model}, is proposed to model the phoneme-level prosody. During the training stage, the prosodic features are represented by pitch, duration, and energy, all of which are at the phoneme level. Meanwhile, a text-based prosody predictor is optimized with the input of text encoder's output and the style embedding. During the inference stage, the prosody predictor is to provide the speaking style information for speech synthesis.

\textbf{Prosody predictor}~The structure of the prosody predictor is shown in Fig.~\ref{fig:pred}. It consists of five one-dimension convolutional layers and one linear transformation layer. Each of the convolutional layer is followed by layer normalization, ReLu activation function, and dropout. Considering the temporal nature of prosodic sequences, a position-coding vector is added to the input. To optimize the prosody predictor, the L1 loss is used to calculate the deviation between the predicted prosody and the ground-truth prosody features.

~\textbf{Multi-scale prosody encoder}~The speaking style of human speech has rich and subtle changes even within the same utterance. These changes are generally reflected in different scales. To obtain better representation from prosodic features, a multi-scale encoder as shown in Fig.~\ref{fig:enc}, is proposed in our framework.
The input prosody features are first convolved with a one-dimensional convolution filter bank $\mathbf{F}=\left\{\mathbf{f}_{1}, \ldots, \mathbf{f}_{m}\right\}$ where $\mathbf{f}_{i}$ has a width of $i$. In practice, $m$ is 8 in the proposed model.
The outputs of the convolution groups are stacked together, and the processed sequence is further passed to the maximum pooling layer and 1-D convolution layer. Then we use a layer of bidirectional LSTM (BLSTM) to extract the forward and backward sequence features. With this multi-scale modeling method, we can explicitly obtain the local and contextual features from the prosodic components. 

\subsection{Style control}
Since the proposed method is based on explicit prosody features, it allows us to control the prosody feature by adjusting its value. Specifically, by multiplying or dividing the prosody features by a scale, we can flexibly control the prosody of the synthesized speech to further enhance the expressiveness of synthesized speech.

\subsection{Training and generation}
The training loss of the proposed model is shown in
\begin{equation}\label{eq1}
\mathcal{L}=\mathcal{L}_{\text {taco}}+\mathcal{L}_{\text {prosody}},
\end{equation}
where $\mathcal{L}_{\text {taco}}$ is the loss function of Tacotron2~\cite{shen2018natural}, and $\mathcal{L}_{\text {prosody}}$ is prosody loss. The training and inference stages are illustrated in Fig.~\ref{fig:model}. At the training stage, ground-truth prosody features are used as input of the prosody encoder. At the inference stage, the prosody features are predicted based on the input text and style id.

\section{Experiments}
\label{sec:typestyle}

\subsection{Experimental setup}

\subsubsection{Database}

To evaluate the performance of the proposed method in the SRM2TTS task, an internal Mandarin multi-speaker corpus, in which each speaker has a unique speaking style, is employed in the experiments. There are a total of six speakers, each with their own unique style, including reading, radio anchor, story telling, customer service (CS), poetry and game character. 
Compared with the first four speaking styles, the latter two have stronger expressiveness, which are recorded by a child and a game character respectively.
The total duration is 20 hours, and all recordings are down-sampled to 16kHz. Ten sentences of each speaker are randomly selected as the test set for subjective evaluation. 

\subsubsection{Evaluation metrics}

In the SRM2TTS task, a good model should has the ability to produce synthetic speech with the expected speaking style and timbre. Therefore, the synthesized results are evaluated in terms of style similarity and speaker similarity.

\textbf{Style similarity}:
The style similarity is to compare the similarity between the expected speaking style of natural speech and that of synthetic speech. Here, a Mean Opinion Score (MOS) evaluation with the human rating experiment is conducted to evaluate this similarity.
Among the speakers from the adopted database, the speaker with the reading style (DB1\footnote{The dataset is available at \url{http://www.data-baker.com/hc_znv_1.html}}) is a public database. Therefore, in the evaluation, DB1 is adopted as the target timbre to express different speaking styles.
Twenty (gender-balanced) native Mandarin listeners are invited to participate in the evaluation.

\textbf{Speaker similarity}:
The speaker similarity is to compare the similarity between the expected timbre of natural speech and that of synthetic speech. Similar to the evaluation of style similarity, MOS evaluation is conducted in the subjective test. 

\subsection{Comparison with other methods}
To evaluate the performance of the proposed model on the SRM2TTS task, two state-of-the-art style transfer methods, i.e., Multi-R~\cite{bian2019multi} and PB~\cite{pan2021cross}, are compared in this work. Multi-R~\cite{bian2019multi} is a Tacotron-based method with multi-reference to transfer the prosody. PB~\cite{pan2021cross} is a cross-speaker style transfer model based on prosodic bottleneck. For the fair comparison, the compared Multi-R and PB take the same Tacotron backbone as our proposed model.

The MOS evaluations in terms of style similarity and speaker similarity are shown in Table~\ref{tab:transfer}. As can be seen from the table, our model achieves the best performance in terms of all style categories. Note that the reference-based method Multi-R obtains the lowest MOS scores with all speaking styles. This is mainly because that when each speaker has a unique speaking style this reference-based method is hard to decouple the timbre and style of the speaker. Therefore, when the imitated speaking style is significantly different from the reading style, i.e., game and poetry, the performance of this reference-based method performs much worse. In contrast, the label-based PB and our method achieve better style similarity MOS scores, which is probably caused by that the distinctive speaking style makes it easier for participants to judge, indicating the effectiveness of the label-based method on this SRM2TTS task. Compared with PB, our proposed method achieves 4.5\% relatively higher style similarity MOS averaging all style categories. 

As for the speaker similarity, no obvious MOS difference exists among three models, demonstrating that the style transfer in PB and the proposed method does not bring obvious negative effect to the timbre compared with Multi-R which has very limited style transfer ability. Instead, the proposed method even achieves the best speaker similarity MOS in terms all style categories except for CS and Poetry, indicating the good performance of the proposed method on the SRM2TTS task.

\begin{table}[]
 \caption{Ablation study of different prosody component in terms of style similarity MOS with confidence intervals of 95$\%$. w/o means without.}
 \label{tab:compare_Ablation}
\setlength{\tabcolsep}{0.1mm}
 \centering
 \small
\begin{tabular}{l|l|l|l|l|l}
\toprule
\multicolumn{1}{c|}{Method} & \multicolumn{1}{c}{Proposed} & \multicolumn{1}{c}{w/o energy} & \multicolumn{1}{c}{w/o duration} & \multicolumn{1}{c}{w/o pitch} & \multicolumn{1}{c}{w/o all} \\ \midrule
Story   & \bf{3.76}$\pm$\bf{0.049} & 3.74$\pm$0.050   & 3.55$\pm$0.053 & 3.65$\pm$0.065 & 3.41$\pm$0.044 \\
Anchor    & \bf{3.85}$\pm$\bf{0.042}  & 3.76$\pm$0.048 & 3.67$\pm$0.063 & 3.73$\pm$0.049 & 3.37$\pm$0.045 \\
CS     & \bf{3.87}$\pm$\bf{0.054} & 3.80$\pm$0.049 & 3.44$\pm$0.054 & 3.66$\pm$0.057  & 3.22$\pm$0.042 \\
Poetry & \bf{3.97}$\pm$\bf{0.045} & 3.81$\pm$0.050 & 3.27$\pm$0.044 & 3.59$\pm$0.062& 2.93$\pm$0.039 \\
Game      & \bf{3.91}$\pm$\bf{0.041} & 3.83$\pm$0.044 & 3.13$\pm$0.043 & 3.49$\pm$0.041 & 2.77$\pm$0.036 \\ \midrule
Overall     & \bf{3.87}$\pm$\bf{0.021} & 3.79$\pm$0.022 & 3.42$\pm$0.021 & 3.63$\pm$0.021 & 3.14$\pm$0.020   \\ 
  \bottomrule
\end{tabular}
\vspace{-0.5cm} 
\end{table}

\subsection{Ablation study}
The prosody prediction module plays an important role to realize the SRM2TTS task. In the prosody prediction module, several prosody components, including pitch, energy, and duration, are considered in our method. To show the effectiveness of each component in the SRM2TTS, an ablation study is performed by comparing the proposed method with several variants that achieved by dropping one or all prosody components. The results are shown in Table~\ref{tab:compare_Ablation}, in which the style similarity is evaluated by the MOS score. Note that, when all of these three prosodic components are removed, which is referred to as \textit{w/o all} in Table~\ref{tab:compare_Ablation}, the model degenerates into a general multi-speaker model. Due to that the human rating experiments in Table~\ref{tab:compare_Ablation} and Table~\ref{tab:transfer} are performed in two individual groups, the MOS scores of the proposed method in these two tables are slightly different.

As can be seen from this table, dropping any prosodic component would result in significant performance drop in terms of the style similarity. Specifically, the dropping of the duration brings the biggest drop, in which the style similarity MOS is 11.7\% relatively lower than the proposed method. When no prosody component is adopted, i.e., \textit{w/o all}, this model is unable to perform the style transfer task. Instead, it is just a multi-speaker TTS model that can only produce synthetic speech with the timbre and style that belongs to the same speaker in the corpus. All of these results indicates the importance of each prosody component in our prosody modeling module. In addition to the effect on the style similarity, those prosody components also show important roles in the manual control of prosody in synthetic speech, which will be demonstrated in Section~\ref{sc:control}.



\begin{figure}[!htbp]
    \centering
    \includegraphics[scale=0.41]{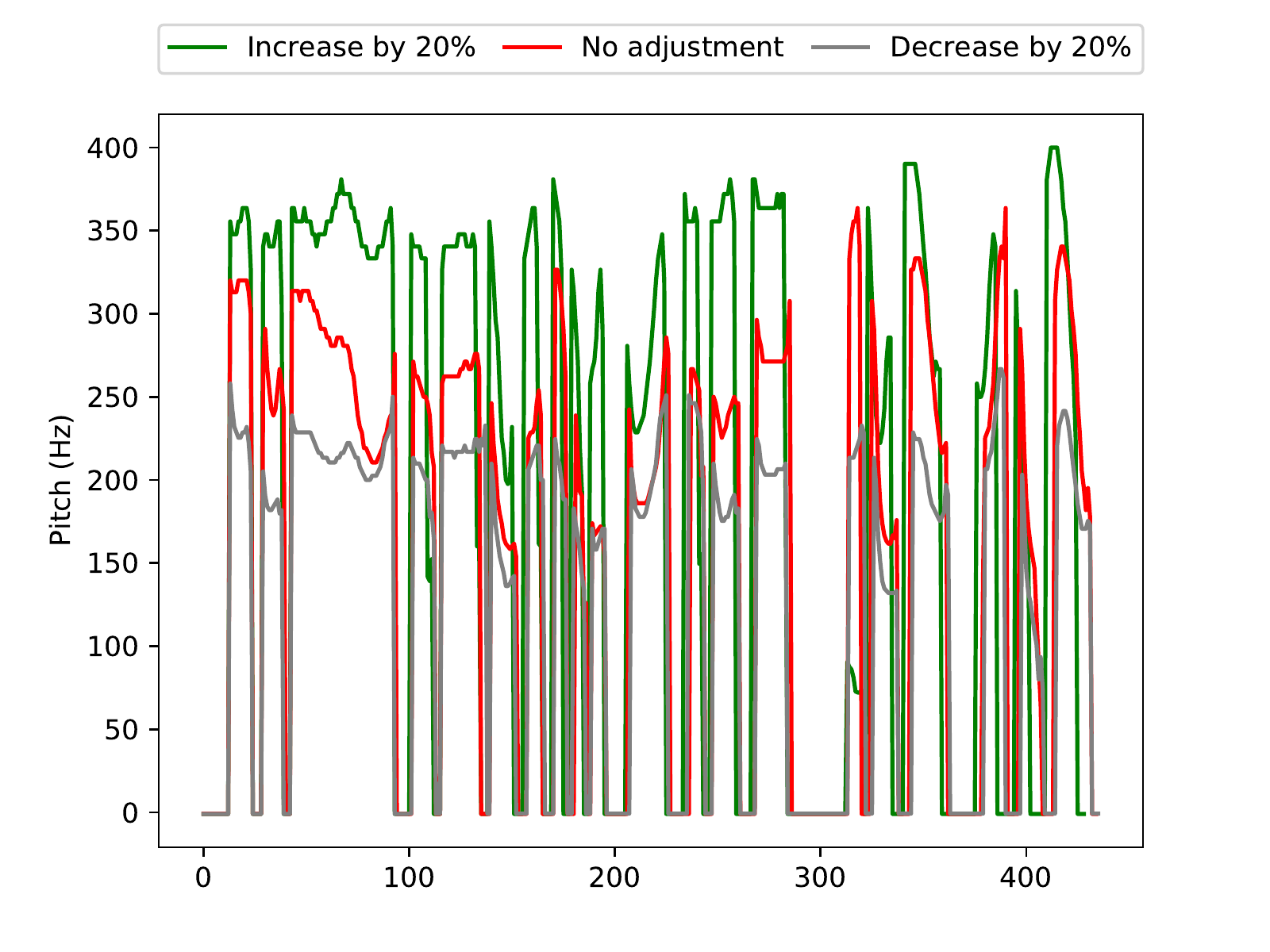}
    \caption{Pitch contours of synthesized speech with pitch component increases by 20$\%$, does not adjust, decreases by 20$\%$.}
    \label{fig:modelf0} 
\end{figure} 

\begin{figure}[t]
    \centering
    \includegraphics[scale=0.41]{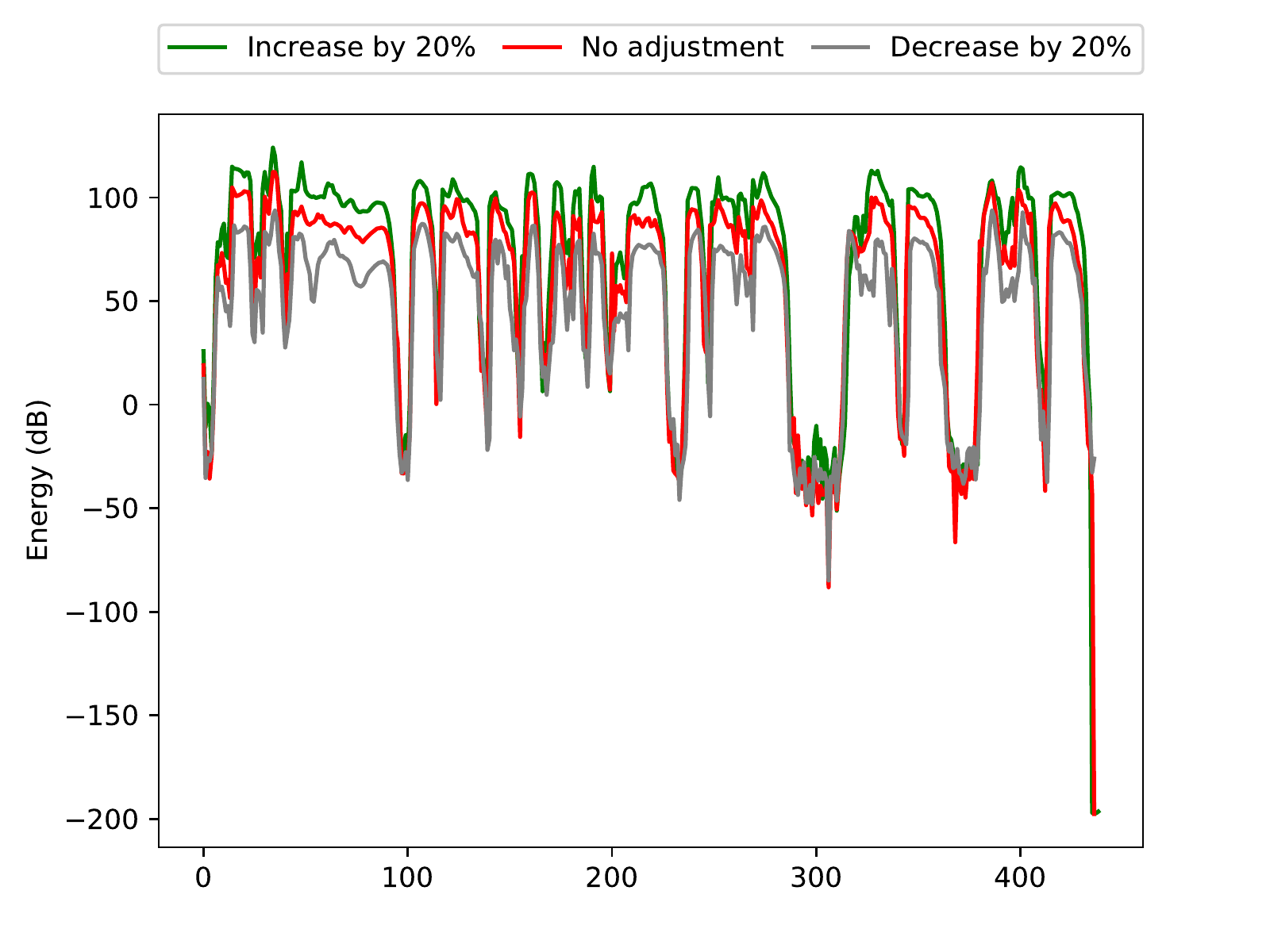}
    \caption{Energy contours of synthesized speech with energy component increases by 20$\%$, does not adjust, decreases by 20$\%$.}
    \label{fig:modelenergy} 
\end{figure} 

\begin{figure}[t]
    \centering
    \includegraphics[scale=0.42]{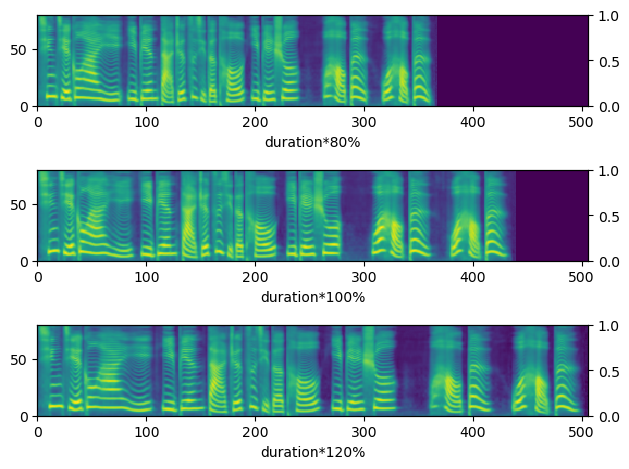}
    \caption{Mel spectrums of synthesized speech with duration component increases by 20$\%$, does not adjust, decreases by 20$\%$.}
    \label{fig:modelduraion} 
\end{figure} 
\vspace{-0.7cm}
\subsection{Style control}
\label{sc:control}
Since we explicitly use the prosody features, i.e., pitch, duration, and energy in the prosody prediction module, we can easily control the prosody by adjusting the prosody features. For instance, we can simply multiply the duration by a scale to control the speaking rate. 

Figs.~\ref{fig:modelf0}-\ref{fig:modelduraion} present different pitch, energy, and Mel spectrums of synthesized speech by adjusting the pitch, energy, and duration respectively. As can be seen, the adjusting of the prosody features can exactly control the corresponding prosody of synthesized speech, indicating that our prosody encoder can model the explicit independent prosody components in the final synthesized speech.
Even though the larger scale means the greater change to the corresponding prosodic component, the scale cannot be infinite. For instance, too short duration or too small energy would affect intelligibility.
In the experiment, we found that the pitch and energy can be effectively controlled within a scale of 20$\%$, and the duration can be successfully controlled within a scale of 50$\%$. Synthesized samples are presented at the demo page\footnote{Audio samples can be found on the project page \url{https://qicongxie.github.io/SRM2TTS}}, and we encourage readers to listen to them.

\vspace{-0.2cm}
\section{conclusion}
\label{sec:refs}
In this paper, a general stylized speech synthesis task is proposed. This task, referred to as SRM2TTS, aims to produce expressive synthetic speech by combining any speaking style of one speaker with a timbre of another speaker. Compared with existing style transfer task, this proposed task is more general and practical as it can bypass the dependency on a source speaker who has to record all expected speaking styles. Therefore, the realization of this task is promising for many application cases. 

With the aim to achieve this SRM2TTS task, a novel style modeling method based on explicit prosody features is proposed. This proposed method is based on the backbone of Tacotron2 and with a fine-grained text-based prosody prediction module and a speaker controller. Extensive experiments have shown that the proposed method can successfully express the style of one speaker with the timbre of another speaker. Furthermore, the explicit use of the prosody feature in the prosody prediction module allows us to control the prosody manually, which can produce more diverse expressive synthetic speech.

\vfill\pagebreak

\bibliographystyle{IEEEbib}
\bibliography{strings,refs}

\end{document}